\documentclass{emulateapj}

\usepackage{amssymb}
\usepackage{amsmath}
\usepackage{epstopdf}
\usepackage{natbib}
\usepackage[colorlinks, citecolor=blue]{hyperref}
\usepackage{hyperref}
\usepackage[all]{hypcap}

\begin{document}

\title{Modeling the Light Curves of the Luminous Type Ic Supernova 2007D}
\author{Shan-Qin Wang\altaffilmark{1,2,3,4}, Zach Cano\altaffilmark{5,6}, Long Li\altaffilmark{2},
Liang-Duan Liu\altaffilmark{1,3,7}, Ling-Jun Wang\altaffilmark{8},\\ WeiKang Zheng\altaffilmark{4},
Zi-Gao Dai\altaffilmark{1,3}, En-Wei Liang\altaffilmark{2}, and Alexei V. Filippenko\altaffilmark{4,9}}

\begin{abstract}

SN~2007D is a nearby (redshift $z = 0.023146$), luminous Type Ic supernova (SN)
having a narrow light curve (LC) and high peak luminosity. Previous research based
on the assumption that it was powered by the $^{56}$Ni cascade decay suggested
that the inferred $^{56}$Ni mass and the ejecta mass are $\sim 1.5$\,M$_{\odot}$
and $\sim 3.5$\,M$_{\odot}$, respectively. In this paper, we employ some multiband
LC models to model the $R$-band LC and the color ($V-R$) evolution of
SN~2007D to investigate the possible energy sources powering them. We
find that the pure $^{56}$Ni model is disfavored; the multiband LCs
of SN~2007D can be reproduced by a magnetar whose initial rotational
period $P_{0}$ and magnetic field strength $B_p$ are $7.28_{-0.21}^{+0.21}$
(or $9.00_{-0.42}^{+0.32}$) ms and $3.10_{-0.35}^{+0.36}\times 10^{14}$
(or $2.81_{-0.44}^{+0.43}\times 10^{14}$) G, respectively.
By comparing the spectrum of SN~2007D with that of some superluminous SNe (SLSNe),
we find that it might be a luminous SN like several luminous ``gap-filler"
optical transients that bridge ordinary and SLSNe, rather than a genuine
SLSN.

\end{abstract}


\affil{\altaffilmark{1}School of Astronomy and Space Science, Nanjing
University, Nanjing 210093, China; dzg@nju.edu.cn}
\affil{\altaffilmark{2}Guangxi Key Laboratory for Relativistic Astrophysics,
School of Physical Science and Technology, Guangxi University, Nanning 530004,
China; shanqinwang@gxu.edu.cn}
\affil{\altaffilmark{3}Key Laboratory of Modern Astronomy and Astrophysics
(Nanjing University), Ministry of Education, China}
\affil{\altaffilmark{4}Department of Astronomy, University of California,
Berkeley, CA 94720-3411, USA}
\affil{\altaffilmark{5}Instituto de Astrof\'isica de Andaluc\'ia (IAA-CSIC),
Glorieta de la Astronom\'ia s/n, E-18008, Granada, Spain}
\affil{\altaffilmark{6}Juan de la Cierva Fellow}
\affil{\altaffilmark{7}Department of Physics and Astronomy,
University of Nevada, Las Vegas, NV 89154, USA}
\affil{\altaffilmark{8}Astroparticle Physics, Institute of High Energy Physics,
Chinese Academy of Sciences, Beijing 100049, China}
\affil{\altaffilmark{9}Miller Senior Fellow, Miller Institute for Basic
Research in Science, University of California, Berkeley, CA 94720, USA}

\section{Introduction}

\label{sec:Intro}

As a subclass of core-collapse supernovae (CCSNe), Type Ic SNe (SNe~Ic) have long been
believed to be the results of explosions of massive stars that had lost all of their
hydrogen and all (or almost all) of their helium envelopes, thereby showing
no hydrogen and helium absorption lines (see \citealt{Fil1997,Mat2011,Gal2017} for
reviews). The light curves (LCs), spectra, and physical parameters of SNe~Ic are rather
heterogeneous. According to their peak luminosities they can be classified into three
subclasses: ordinary SNe~Ic, luminous SNe~Ic, and superluminous SNe~Ic ({SLSNe~Ic;
\citealt{Qui2011,Gal2012,Gal2018}).\footnote{See, e.g., Figure 13 of \citet{Nich2015} and
Figure 3 of \citet{DeCia2018}. \citet{DeCia2018} show that there is a continuous
luminosity function from faint SNe~Ic to SLSNe-I. We call SNe that are dimmer than
SLSNe but brighter than canonical SNe Ia ``luminous SNe"; they are similar to the
luminous optical transients presented by \citet{Arc2016}.}

Based on their spectra around peak brightness, SNe~Ic can be divided into normal
SNe~Ic and ``broad-lined SNe~Ic (SNe~Ic-BL)" \citep{Woo2006}.
And, according to their kinetic energy ($E_{\rm K}$), they can be
split into normal SNe~Ic ($E_{\rm K} \lesssim 2\times10^{51}$ erg)
and ``hypernovae" ($E_{\rm K} \gtrsim 2\times10^{51}$ erg; \citealt{Iwa1998}).
A minority of SNe~Ic-BL are associated with gamma-ray bursts (GRBs) or
X-ray flashes (XRFs) and were called ``GRB-SNe"
(see \citealt{Woo2006,Hjo2012,Cano2017}, and references therein).

Study of the energy sources of SNe~Ic-BL and SLSNe~I/Ic is a very important
part of time-domain astronomy. The LCs of normal SNe~Ic can be explained by
the $^{56}$Ni cascade decay model ($^{56}$Ni model for short; \citealt{Col1969,Col1980,Arn1982}),
while the energy sources of luminous SNe and SLSNe are still being debated:
they cannot be explained by the $^{56}$Ni model (e.g., \citealt{Qui2011,Gal2012,Inse2013}),
so instead researchers often invoke the magnetar model
\citep{Mae2007,Kas2010,Woos2010,Cha2012,Cha2013,Des2012b,Inse2013,Chen2015,Wang2015a,Wang2016b,Dai2016},
involving nascent highly magnetized neutron stars (magnetic strength $B_{p}
\approx 10^{13}$--$10^{15}$ G)\footnote{It was suggested that a magnetar with
$B_{p} \approx 10^{16}$ G can power SNe~Ic-BL \citep{Wang2016a,Wang2017a,Wang2017b,Chen2017}.},
and the circumstellar interaction model \citep{Che1982,Che1994,Chu1994,Gin2012,Cha2012,Cha2013,Liu2018},
in which ejecta kinetic energy is converted to radiation.

In this paper, we study the very nearby Type Ic SN~2007D.
The luminosity distance $D_L$ derived from the Tully-Fisher
relation and the redshift $z$ of the host galaxy of SN~2007D (UGC~2653) are
$106_{-8.5}^{+2}$\,Mpc (from NED)\footnote{http://ned.ipac.caltech.edu/cgi-bin/nDistance?name=UGC+02653 .}
and $0.023146\pm0.000017$ (recession velocity $6939\pm5$ km s$^{-1}$; \citealt{Weg1993}),
respectively. The photospheric velocity ($v_\mathrm{ph}$) of SN~2007D
inferred from the Fe\,{\sc ii}$\lambda$5169 absorption line about 8 days before
$V$-band maximum brightness is $\sim 13,350 \pm 4000$ km s$^{-1}$ \citep{Mod2014,Mod2016},
smaller than the canonical value of SNe~Ic-BL ($\sim 22,200 \pm 9400$ km s$^{-1}$;
\citealt{Mod2016}) and the average values of SLSNe~I ($\sim 15,000 \pm 2600$ km s$^{-1}$;
\citealt{Liu2017b}) 10 days after peak brightness.

SN~2007D was heavily extinguished by its highly inclined ($\sim 70^{\circ}$;
\citealt{Dro2011}) host galaxy UGC~2653 ($E(B-V)_\mathrm{host} = 0.91 \pm 0.13$
mag; \citealt{Dro2011}) and the Milky Way ($E(B-V)_\mathrm{Gal} = 0.335$ mag;
\citealt{Sch1998}). By performing the extinction correction, \citet{Dro2011}
found that the $R$-band and $V$-band peak absolute magnitudes
($M_{R,{\mathrm{peak}}}$ and $M_{V,{\mathrm{peak}}}$) of SN~2007D are
$\sim -20.65 \pm 0.55$ mag and $< -20.54$ mag, respectively,
significantly brighter than all other SNe~Ibc.\footnote{The average peak absolute
magnitude of two dozen nearby ($D_L \lesssim 60$ Mpc)
SNe~Ibc discovered by the Lick Observatory Supernova Search (LOSS) is $-16.09 \pm 0.23$ mag
(with a 1$\sigma$ dispersion of 1.24 mag; \citealt{Li2011}). The average peak absolute magnitude of
nearby ($D_L \lesssim 150$ Mpc) SNe~Ic and SNe~Ic-BL observed by the Palomar 60-inch telescope (P60)
are $-17.4 \pm 0.4$ mag and $-18.3 \pm 0.6$ mag, respectively \citep{Dro2011}. Among these SNe~Ic
and SNe~Ic-BL, SN~2007D is the most luminous.}
While \citet{Gal2012} suggested that the SLSN threshold can be set at $-21$ mag,
\citet{Qui2018} and \citet{DeCia2018} re-examined the threshold of SLSNe and
suggested it is $\sim -20.5$ mag, as adopted by \citet{Qui2014}.
According to the latter threshold, SN~2007D is a SLSN.
However, the extinction values of the host galaxy of SN~2007D and the
Milky Way are rather uncertain. For example, using the values of
\citet{Sch2011} for the
foreground extinction\footnote{http://irsa.ipac.caltech.edu/applications/DUST/}
(which are roughly 20--30\% lower than those of \citealt{Sch1998})
and the $K$-corrected $V$-band LC of SN~2007D, we find a peak absolute
magnitude $M_\mathrm{V,peak}$ of only $\sim -20.06$ mag\footnote{This arises from
a peak apparent magnitude of $m_\mathrm{V,peak} = 15.06 \pm 0.36$,
which includes a foreground extinction of 0.79 mag and host extinction of 2.50 mag,
and the Tully-Fisher distance modulus on the NED website
(http://ned.ipac.caltech.edu/cgi-bin/nDistance?name=UGC+02653) of $35.12 \pm 0.47$ mag.},
$\sim 0.48$ mag dimmer than the value inferred by \citet{Dro2011} ($< -20.54$ mag).
In this case, SN~2007D is a luminous SN whose peak luninosity is between that
of ordinary SNe and SLSNe (see, e.g., \citealt{Arc2016}). We call these two different LCs
``Case A" and ``Case B" throughout this paper.

The energy source of SN~2007D has not yet been definitively determined.
By assuming that the luminosity evolution of SN~2007D was powered by $^{56}$Ni decay
and supposing that the ejecta velocity is $\sim 2\times 10^{9}$ cm s$^{-1}$,
\citet{Dro2011} inferred that the mass of $^{56}$Ni synthesized in the explosion
and the value of $(M_{\mathrm{ej}}/{\rm M}_\odot)^{3/4}(E_{\mathrm{K}}/10^{51} \mathrm{erg})^{-1/4}$
are $\sim 1.5 \pm 0.5~$M$_\odot$ and $\sim 1.5_{-0.5}^{+0.8}~$M$_\odot$, respectively (see
Table 6 of \citealt{Dro2011}). Supposing $v_\mathrm{sc} \approx 2\times 10^{9}$ cm s$^{-1}$
for the scale velocity of the ejecta
and solving the equation $(M_{\mathrm{ej}}/\mathrm{M}_\odot)^{3/4}(E_{\mathrm{K}}/10^{51}
\mathrm{erg})^{-1/4} = 1.5_{-0.5}^{+0.8}$, however, we find that the mass of
the ejecta $M_{\mathrm{ej}} = 3.5_{-1.95}^{+4.7}~$M$_\odot$.
Then the ratio of the $^{56}$Ni mass to the ejecta mass ($M_{\mathrm{Ni}}/M_{\mathrm{ej}}$)
is $\sim 0.43_{-0.31}^{+0.86}$, significantly larger than
the upper limit ($\sim 0.20$) determined by numerical simulations \citep{Ume2008},
suggesting that the photometric evolution of SN~2007D cannot be explained by the
$^{56}$Ni model. Therefore, the question of the energy source of SN~2007D deserves
detailed study. In fact, \citet{Gal2012} had discussed SN~2007D and SN~2010ay
as ``transitional'' events between SLSNe-I and SNe~Ic
and suggested that a ``central engine"
may power their large observed peak luminosities. However, no quantitative
research on this idea has been performed to date.

In this paper, we investigate in detail the energy-source mechanisms powering the luminosity
evolution of SN~2007D. In Section \ref{sec:fit}, we employ the $^{56}$Ni model, the magnetar model,
as well as the magnetar+$^{56}$Ni model to fit the $R$-band LC and the $V-R$ color evolution
of SN~2007D. Discussion and conclusions are presented in Sections \ref{sec:dis} and \ref{sec:con},
respectively.

\section{Modeling the Multiband LCs of SN~2007D}

\label{sec:fit}

In this Section, we employ semianalytic models to fit the $R$-band LC
and the $V-R$ color evolution of SN~2007D.\footnote{The $R$, $V$, and $V-R$ LCs are
presented by \citet{Dro2011}. By fitting two of these three LCs, the remaining
one is also determined. We choose to fit the $R$ and $V-R$ LCs.}
To fit these LCs, we neglect the dilution effect (e.g., \citealt{Des2012a})
of the ejecta and assume that the SN radiation is black-body emission:
$F(\nu,t)= (2{\pi}h{\nu}^3/c^2)(e^{\frac{h{\nu}}{k_{\rm b}T(t)}}-1)^{-1}(R^2/D_L^2)$, where
$T(t)= (L(t)/4\pi\sigma(v_{\mathrm{sc}}t)^2)^{1/4}$ is the black-body temperature and
$L(t)$ is the bolometric luminosity of a SN. Using the Vega magnitude system
($\mathrm{mag}(\nu,t) = -2.5\,\mathrm{log}_{10}F(\nu,t)-48.598-zp(f_{\nu})$) and
Table A2 of \citet{Bes1998}, we can convert
the fluxes to magnitudes.\footnote{In Table A2 of \citet{Bes1998}, note
that ``$zp(f_{\lambda}$)" (in the fourth line) and ``$zp(f_{\nu}$)"
(in the fifth line) must be exchanged.}
Hence, our semianalytic models should simultaneously reproduce the bolometric LC, the
temperature evolution, and the multiband LCs of SN~2007D.
In adopting a simple black-body model, we neglect the blue-ultraviolet (UV)
suppression which yields a dimmer blue-UV luminosity and a brighter
optical luminosity. To get the best-fit parameters and the range,
we adopt the Markov Chain Monte Carlo (MCMC) method.

\subsection{The $^{56}$Ni-Only Model}

We first employ a semianalytic $^{56}$Ni model to fit the $R$ and
$V-R$ LCs. The LCs reproduced by this model are determined by the optical opacity $\kappa$,
the ejecta mass $M_{\mathrm{ej}}$, the initial scale velocity of the ejecta $v_{\mathrm{sc0}}$,
the $^{56}$Ni mass $M_{\mathrm{Ni}}$, the gamma-ray opacity of $^{56}$Ni decay
photons $\kappa_{\gamma,\mathrm{Ni}}$, and the moment of explosion $t_\mathrm{expl}$.
We suppose that the initial kinetic energy of the ejecta
($E_{\mathrm{K0}} = 0.3\,M_{\mathrm{ej}}v_{\mathrm{sc0}}^{2}$) is provided by
the neutrino-driven mechanism. Then the upper limit of $E_{\mathrm{K0}}$ is set to
be $2.5 \times 10^{51}$ erg since the upper limit of $E_{\mathrm{K0}}$
provided by the neutrino-driven mechanism is (2.0--2.5) $\times 10^{51}$ erg; \citealt{Jan2016}.
The upper limit of $v_{\mathrm{sc0}}$ is adopted to be $\sim 16,000$ km s$^{-1}$.
Without this constraint, MCMC would favor a $v_{\mathrm{sc0}}$ value that yields
a photospheric velocity significantly larger than the observed one
($\sim 13,350 \pm 4000$ km s$^{-1}$) since there is only one velocity point.

The theoretical $^{56}$Ni-powered $R$ and $V-R$ LCs are shown in
Figure \ref{fig:2007D-Ni}. The parameters of the $^{56}$Ni model are listed
in Table \ref{tab:para}. To match the post-peak $R$-band LC, the value
of $\kappa_{\gamma,\mathrm{Ni}}$ must be $1.12_{-0.86}^{+4.01}$ cm$^{2}$ g$^{-1}$,
larger than the canonical value of 0.027 cm$^{2}$ g$^{-1}$ (e.g., \citealt{Cap1997,Maz2000,Mae2003}).

For Case A, the inferred $^{56}$Ni mass is $\sim 2.66_{-0.15}^{+0.17}$\,M$_\odot$.
This value is significantly larger than that ($\sim 1.5$\,M$_\odot$) derived from a relation
linking the $R$-band peak magnitude $M_{R,\mathrm{peak}}$ and the $^{56}$Ni mass yield used by
\citet{Dro2011}. This is because higher peak luminosity and temperature result in
a bluer photosphere when the SN peaks and the ratio of the UV flux to
the $R$-band flux is larger than that of the normal SNe~Ibc, and more $^{56}$Ni
is needed for powering the SN peak. As shown in Table \ref{tab:para}, the derived
ejecta mass is $1.39_{-0.33}^{+0.19}$\,M$_\odot$, smaller than the mass of $^{56}$Ni.
For Case B, the inferred values of the ejecta mass and $^{56}$Ni mass
are $1.45_{-0.32}^{+0.17}$\,M$_\odot$ and $1.61_{-0.07}^{+0.08}$\,M$_\odot$,
respectively. The $^{56}$Ni mass is also larger than the ejecta mass.

We note that the value of $\kappa$ can vary from 0.06 to 0.20 cm$^2$ g$^{-1}$
(see the references listed by \citealt{Wang2017c}) and was fixed here to
be 0.07 cm$^{2}$ g$^{-1}$. A larger (smaller) value would result in a
smaller (larger) value of $M_{\mathrm{ej}}$ (see, e.g.,
\citealt{Wang2015b,Nagy2016,Wang2017c}). Nevertheless, the inferred ratio of
the $^{56}$Ni mass to the ejecta mass would still be larger than 1.36 (for Case A)
or 0.90 (for Case B) even if $\kappa = 0.06$ cm$^{2}$ g$^{-1}$. 

These results indicate that the $^{56}$Ni model cannot explain the multiband LCs
of SN~2007D and that there might be other energy sources involved, because
the ratio of the $^{56}$Ni mass to the ejecta mass cannot be larger than
$\sim 0.20$ \citep{Ume2008}.

\subsection{The Magnetar Model}
\label{subsec:fit2}

Since the modeling disfavors the $^{56}$Ni-only model,
alternative models must be considered. Here we use the magnetar model
to fit the $R-$ band LC and the color evolution of SN~2007D. The free
parameters of the magnetar model are
$\kappa$, $M_{\mathrm{ej}}$, $v_{\mathrm{sc0}}$, the magnetic strength
$B_{p}$, the magnetar's initial rotational period $P_{0}$,
the gamma-ray opacity of magnetar photons $\kappa_{\gamma,\mathrm{mag}}$,
 and $t_\mathrm{expl}$.

The $R$ and $V-R$ LCs reproduced by the magnetar model
are shown in Figure \ref{fig:2007D-mag} and the corresponding parameters
are listed in Table \ref{tab:para}. We find that a magnetar with
$P_0 \approx 7.28_{-0.21}^{+0.21}$\,ms
(or $9.00_{-0.42}^{+0.32}$\,ms for Case B)
and $B_p \approx 3.10_{-0.35}^{+0.36}\times10^{14}$\,G
(or $2.81_{-0.44}^{+0.43}\times10^{14}$\,G for Case B) can power
the multiband LCs of SN~2007D.

\subsection{The Magnetar Plus $^{56}$Ni Model}
\label{subsec:fit3}

It has been proposed that $\lesssim 0.2$\,M$_\odot$ of $^{56}$Ni can be
synthesized by an energetic SN explosion \citep{Nom2013}.
We employ the magnetar plus $^{56}$Ni model whose free parameters are
$\kappa$, $M_{\mathrm{ej}}$, $v_{\mathrm{sc0}}$, $B_{p}$, $P_{0}$,
$\kappa_{\gamma,\mathrm{mag}}$, $M_{\mathrm{Ni}}$,
$\kappa_{\gamma,\mathrm{Ni}}$, and $t_\mathrm{expl}$.
It can be expected that the contribution of such a small amount of
$^{56}$Ni is substantially less than that of a magnetar. Therefore,
the LCs reproduced by the magnetar and the magnetar plus
$\lesssim 0.2$\,M$_\odot$ of $^{56}$Ni models cannot be distinguished
if we tune the parameters. We add $0.2$\,M$_\odot$ of $^{56}$Ni
(see also \citealt{Met2015,Bers2016} for SN~2011kl) and fit the
LCs.

The LCs produced by such a magnetar ($P_0 \approx 7.43_{-0.21}^{+0.22}$\,ms for
Case A, $9.02_{-0.57}^{+0.44}$\,ms for Case B;
$B_p \approx 3.04_{-0.37}^{+0.37}\times10^{14}$\,G for Case A,
$2.49_{-0.46}^{+0.49}\times10^{14}$\,G for Case B) 
plus $0.2$\,M$_\odot$ of $^{56}$Ni
as well as the LCs powered by $0.2$\,M$_\odot$ of $^{56}$Ni
are plotted in Figure \ref{fig:2007D-magni}, and
the corresponding parameters are listed in Table \ref{tab:para}.
While the photometric evolution of SN~2007D can also be explained by the
magnetar plus $^{56}$Ni model, the contribution of $^{56}$Ni
can be neglected.

\section{Discussion}

\label{sec:dis}

\subsection{Bolometric LC and the Temperature Evolution of SN~2007D}

In Section \ref{sec:fit}, we used several models to fit the $R$ and $V-R$ LCs
of SN~2007D.
To obtain more information, we plot the theoretical bolometric LCs and
the temperature evolution; see Figure \ref{fig:2007D-BoloTV}.
The derived temperature of SN~2007D \text{in Case A} is rather high, $>10,000$ K when
$t-t_\mathrm{peak,bol}\leq 10$ days ($t_\mathrm{peak,bol}$ of SN~2007D
is $\sim 10$ days), comparable to that of SLSNe (see, e.g., Figure 5 of \citealt{Inse2013})
and significantly higher than that of ordinary SNe~Ic at
the same epoch ($\lesssim 7,000$ K; \citealt{Liu2017b}).
The derived temperature of SN~2007D at the same epoch in Case B
is 8000--9000~K, between that of SLSNe-I and ordinary SNe~Ic.

We compare the spectrum of SN~2007D with spectra of
three SLSNe-I (LSQ14bdq, SN~2016aj, and SN~2015bn) at the same epoch
(see Figure \ref{fig:spec}), finding that SN~2007D
is redder than these SLSNe. This result indicates
that the temperature of SN~2007D is lower than the temperature of these three SLSNe-I
and that Case B is favored --- that is,
SN~2007D might be a luminous SN~Ic rather than a SLSN-I. 

\subsection{Physical Parameters of the Ejecta of SN~2007D and the Magnetar}

The physical properties of the ejecta of SN~2007D deserve further discussion.
We focus on the properties derived from the magnetar model and the magnetar
plus $^{56}$Ni model since the $^{56}$Ni-only model was disfavored.

The ejecta mass of SN~2007D inferred by the magnetar plus $^{56}$Ni model
is $\sim 1.3\,$M$_\odot$, smaller that the average values of the ejecta of
SNe~Ic and Ic-BL, but at the lower end of the mass distribution
of magnetar-powered SLSNe \citep{Nich2015,Liu2017a,Yu2017,Nich2017}.
The inferred ejecta mass suggests that the progenitor
of SN~2007D might be in a binary system and experienced mass transfer and/or
line-driven wind emission. A low mass results in a rather short rise time
($t_\mathrm{peak,bol} \approx 10$ days),
comparable to that of SN~1994I which is a SN~Ic
(e.g., \citealt{Nom1994,Iwa1994,Fil1995,Sau2006}) and that
of several luminous ``gap-filler" optical transients bridging ordinary SNe and
SLSNe \citep{Arc2016}.

By adopting the equation $\tau_{m}=(2{\kappa}M_{\mathrm{ej}}/{\beta}v_{\mathrm{sc}}c)^{1/2}$
(where $\beta = 13.8$ is a constant; \citealt{Arn1982}),
we conclude that the diffusion timescale $\tau_{m}$ is $\sim$ 8.3 days.
The values of $P_0$ and $B_{p}$ of the magnetar are $\sim 7.4$\,ms (or $\sim 9.0$\,ms for Case B) 
and $3\times 10^{14}$\,G (or $2.5\times 10^{14}$\,G for Case B), respectively.
Hence, the magnetar's initial rotational energy
$E_{\mathrm{rot,0}} \approx 2 \times 10^{52} \left({P_{0}}/{1~\mathrm{ms}}\right)^{-2}$ erg
and spin-down timescale
$\tau_{p} = 5.3\,(B_{p}/10^{14}~{\rm G})^{-2}(P_0/1~{\rm ms})^2~{\rm yr}$
are $\sim 3.65 \times 10^{50}$ (or $\sim 2.47 \times 10^{50}$) erg
(a factor of 5$-$7 smaller than $E_{\mathrm{K0}}$) and 32.3 (or 68.7) days, respectively.

\section{Conclusions}

\label{sec:con}

SN~2007D is a very nearby SN~Ic whose luminosity distance and redshift are
$106_{-8.5}^{+2}$ Mpc and $0.023146 \pm 0.000017$, respectively.
\citet{Dro2011} demonstrated that SN~2007D is a very luminous SN~Ic:
$M_{R,{\mathrm{peak}}} \approx -20.65 \pm 0.55$ mag and
$M_{V,{\mathrm{peak}}} < -20.54$ mag, which are brighter than
the SLSN threshold ($-20.5$ mag) given by \citet{Qui2018} and \citet{DeCia2018},
and inferred that the $^{56}$Ni powering the luminosity evolution of SN~2007D
is $1.5 \pm 0.5 $\,M$_\odot$. Adopting the values of \citet{Sch2011} for the
foreground extinction and the $K$-corrected $V$-band LC of SN~2007D, however, we found
a peak absolute magnitude $M_\mathrm{V,peak}$ of only $\sim -20.06$ mag, $\sim 0.48$
mag dimmer than the LCs of \citet{Dro2011}.

Our simple estimate shows that the ratio of $^{56}$Ni to the ejecta mass
of SN~2007D is unrealistic large ($\sim 0.43_{-0.31}^{+0.86}$). To verify the
validity of the $^{56}$Ni cascade decay model, we use the $^{56}$Ni model to
fit its $R$ and $V-R$ LCs and find that the required $^{56}$Ni mass
($\sim 2.66_{-0.15}^{+0.17}\,$M$_\odot$ for Case A
or $1.61_{-0.07}^{+0.08}\,$M$_\odot$ for Case B) 
is larger than the inferred ejecta mass
($\sim 1.39_{-0.33}^{+0.19}\,$M$_\odot$ for Case A
or $\sim 1.45_{-0.32}^{+0.17}\,$M$_\odot$ for Case B) 
if its multiband LCs were solely powered by $^{56}$Ni, indicating
that the $^{56}$Ni model cannot account for the LCs of SN~2007D.
Alternatively, we employ the magnetar model and
find that the LCs can be fitted and the parameters are reasonable if the
initial period $P_0$ and the magnetic strength $B_p$ of the putative magnetar
are $7.28_{-0.21}^{+0.21}$\,ms (or $9.00_{-0.42}^{+0.32}$\,ms for Case B)
and $3.10_{-0.35}^{+0.36} \times 10^{14}$\,G
(or $2.81_{-0.44}^{+0.43} \times 10^{14}$\,G for Case B), respectively.

By comparing the LCs reproduced by the magnetar model and the magnetar plus
$^{56}$Ni model (the mass of $^{56}$Ni is set to be $0.2$\,M$_\odot$),
we find that the contribution of $^{56}$Ni was significantly lower
than that of the magnetar and can be neglected; it
is very difficult to distinguish between the LCs reproduced by these two models.
Nevertheless, a moderate amount of $^{56}$Ni is needed since the shock launched
from the surface of the proto-magnetar would heat the silicon shell located at
the base of the SN ejecta and $\lesssim 0.2\,$M$_\odot$ of $^{56}$Ni would be synthesized.
According to these results, we suggest that SN~2007D might be powered
by a magnetar or a magnetar plus $\lesssim 0.2\,$M$_\odot$ of $^{56}$Ni.

Adopting the SLSN threshold ($-20.5$ mag) given by \citet{Qui2018} and
\citet{DeCia2018}, and assuming that the peak magnitudes of $R$ and $V$
LCs of SN~2007D are $-20.65 \pm 0.55$ mag and $< -20.54$ mag (respectively),
one can conclude that SN~2007D is a SLSN. If we use the values of \citet{Sch2011}
for the foreground extinction, however, the luminosity of SN~2007D would be
$\sim 0.48$ mag dimmer, and thus only a luminous SN rather than a SLSN.
The spectrum provides additional evidence to discriminate these two possibilities.
We find that the extinction-corrected premaximum spectrum of SN~2007D is redder
than that of three comparison SLSNe-I (LSQ14bdq, SN~2016aj, and SN~2015bn)
at a similar epoch, indicating that the temperature of SN~2007D is lower
than that of these objects. This fact favors the possibility that
SN~2007D is a luminous SN rather than a SLSN.

\acknowledgments
We thank the anonymous referee for constructive suggestions
that led to improvements in our manuscript. 
This work is supported by the National Basic Research Program
(``973" Program) of China (grant 2014CB845800), the National Key Research and Development
Program of China (Grant No. 2017YFA0402600), and the National Natural Science
Foundation of China (grant 11573014). S.Q.W. and L.D.L. are supported by the China
Scholarship Program to conduct research at U.C. Berkeley and UNLV, respectively.
L.J.W. is supported by the National Program on Key Research and Development
Project of China (grant 2016YFA0400801). A.V.F.'s supernova group is grateful for financial
assistance from the Christopher R. Redlich Fund, the TABASGO Foundation,
and the Miller Institute for Basic Research in Science (U.C. Berkeley).
This research has made use of the CfA Supernova Archive which has been
funded in part by the National Science Foundation through grant AST 0907903,
the Weizmann Interactive Supernova Data Repository (WISeREP), and the
Transient Name Server.


\clearpage

\begin{table*}[tbp]
\caption{Parameters of the various models. The uncertainties are 1$\sigma$.}
\label{tab:para}
\begin{center}
{\scriptsize
\begin{tabular}{ccccccccccccc}
\hline\hline
& $\kappa$ & $M_{\mathrm{ej}}$ & $M_{\mathrm{Ni}}$ & $B_p$ & $P_0$ & $v_{\mathrm{sc0}}$ & $\kappa_{\gamma,\mathrm{Ni}}$ & $\kappa_{\gamma,\mathrm{mag}}$ & $t_\mathrm{expl}$$^{\star}$ \\
& (cm$^2$ g$^{-1}$) & (M$_{\odot}$) & (M$_{\odot}$) & ($10^{14}$~G) & (ms) & ($10^9$cm s$^{-1}$) & (cm$^2$ g$^{-1}$) & (cm$^2$ g$^{-1}$) & (days) \\
\hline
\hline
{\bf Case A}\\
\hline
$^{56}$Ni          & 0.07 & $1.39_{-0.33}^{+0.19}$  & $2.66_{-0.15}^{+0.17}$ &  -  &  -  & $1.59_{-0.02}^{+0.01}$ &    $1.12_{-0.86}^{+4.01}$   &     -     & $-$$9.83_{-0.26}^{+0.13}$  \\
magnetar           & 0.07 & $1.23_{-0.40}^{+0.29}$  & 0   & $3.10_{-0.35}^{+0.36}$ & $7.28_{-0.21}^{+0.21}$ & $1.59_{-0.02}^{+0.01}$ &     -    & $5.01_{-4.27}^{+35.73}$ & $-$$9.83_{-0.28}^{+0.13}$  \\
magnetar+$^{56}$Ni & 0.07 & $1.22_{-0.39}^{+0.30}$  & 0.2 & $3.04_{-0.37}^{+0.37}$ & $7.43_{-0.21}^{+0.22}$ & $1.59_{-0.02}^{+0.01}$ &   $0.027$  & $5.01_{-4.30}^{+33.89}$ & $-$$9.84_{-0.26}^{+0.12}$  \\
\hline
\hline
{\bf Case B}\\
\hline
$^{56}$Ni          & 0.07 & $1.45_{-0.32}^{+0.17}$  & $1.61_{-0.07}^{+0.08}$ &  -  &  -  & $1.58_{-0.03}^{+0.02}$ &    $1.55_{-1.12}^{+3.95}$   &     -     & $-$$9.67_{-0.46}^{+0.24}$  \\
magnetar           & 0.07 & $1.25_{-0.40}^{+0.29}$  & 0   & $2.81_{-0.44}^{+0.43}$ & $9.00_{-0.42}^{+0.32}$ & $1.58_{-0.03}^{+0.01}$ &     -    & $5.25_{-4.52}^{+35.49}$ & $-$$9.80_{-0.32}^{+0.15}$  \\
magnetar+$^{56}$Ni & 0.07 & $1.24_{-0.36}^{+0.29}$  & 0.2 & $2.49_{-0.46}^{+0.49}$ & $9.02_{-0.57}^{+0.44}$ & $1.58_{-0.02}^{+0.01}$ &   $0.027$  & $5.75_{-5.05}^{+34.98}$ & $-$$9.80_{-0.32}^{+0.15}$  \\
\hline\hline
\end{tabular}}
\end{center}
\par
{$\star$ The value of $t_\mathrm{expl}$ is with respect to the date of the first $R$-band observation; the lower limit is set to be $-10$ here. \newline}
\end{table*}

\clearpage

\begin{figure}[tbph]
\begin{center}
\includegraphics[width=0.45\textwidth,angle=0]{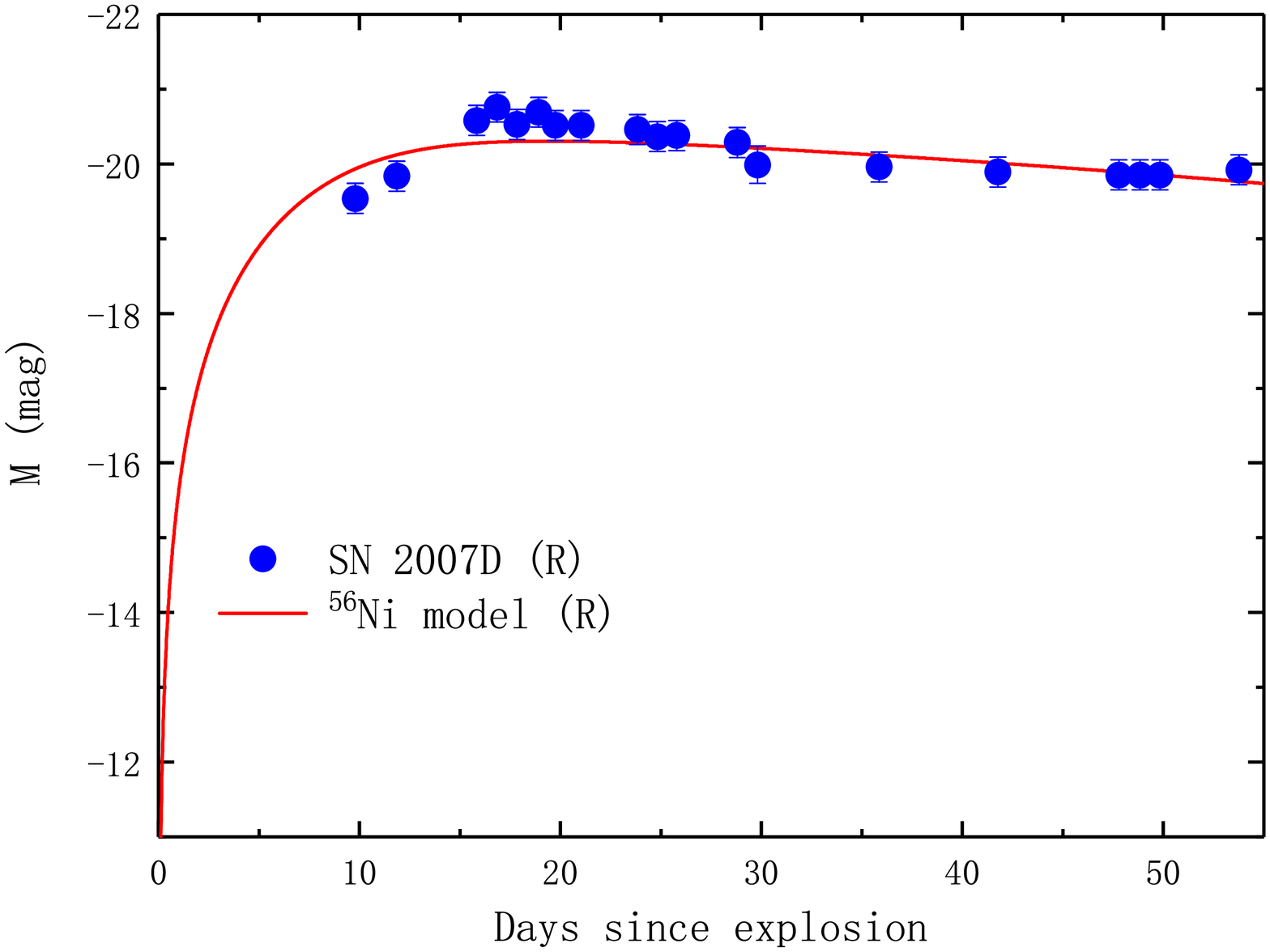}
\includegraphics[width=0.45\textwidth,angle=0]{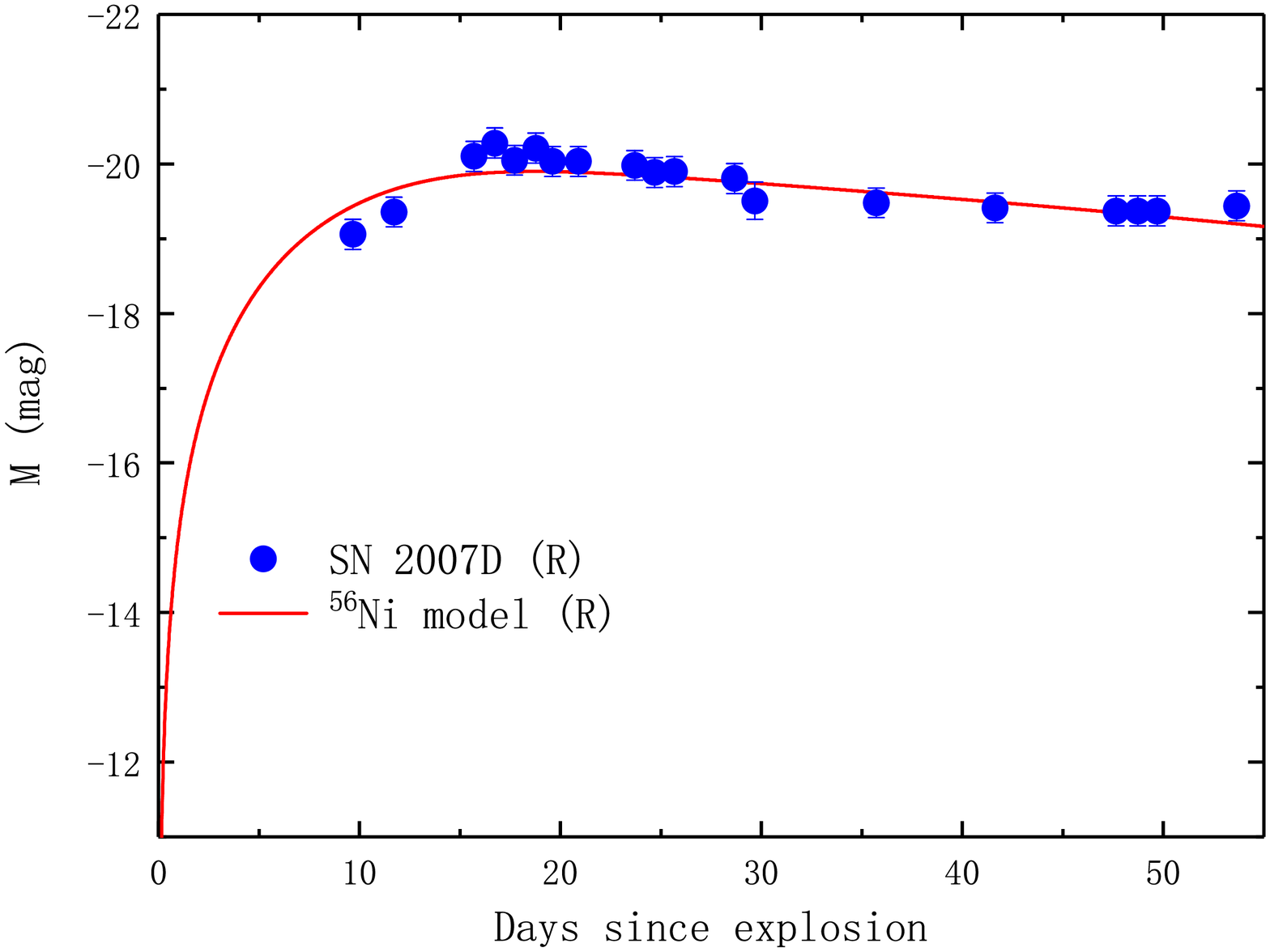}
\includegraphics[width=0.45\textwidth,angle=0]{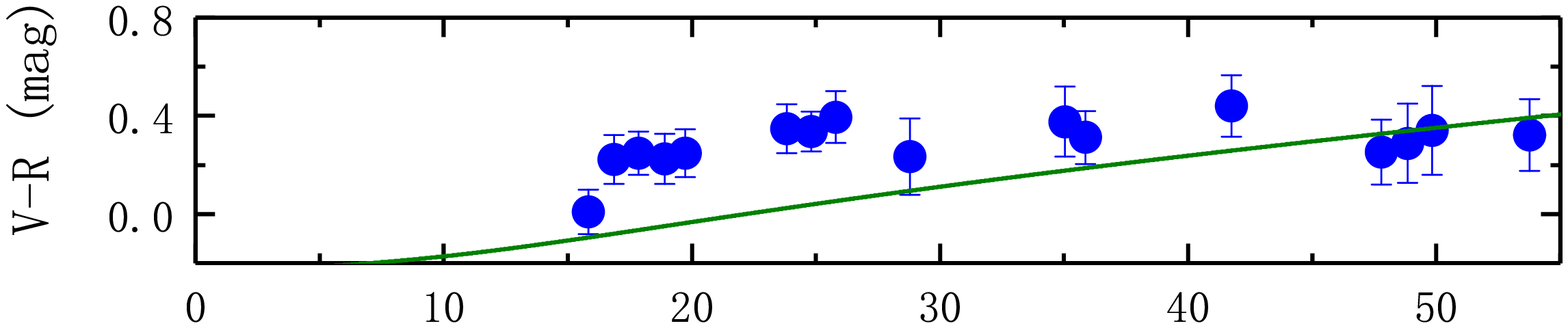}
\includegraphics[width=0.45\textwidth,angle=0]{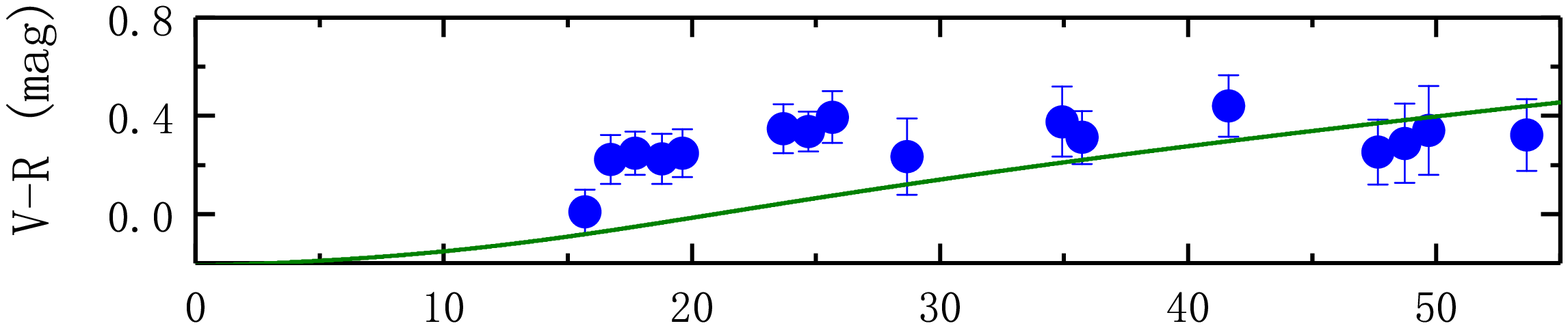}
\end{center}
\caption{The $R$-band LCs (top panels) and the color ($V-R$) evolution
(bottom panels) reproduced by the $^{56}$Ni model for Case A (left) and Case B (right). 
Data for Case A are taken from \citet{Dro2011}. The abscissa
represents time since the explosion in the rest frame.}
\label{fig:2007D-Ni}
\end{figure}

\clearpage

\begin{figure}[tbph]
\begin{center}
\includegraphics[width=0.45\textwidth,angle=0]{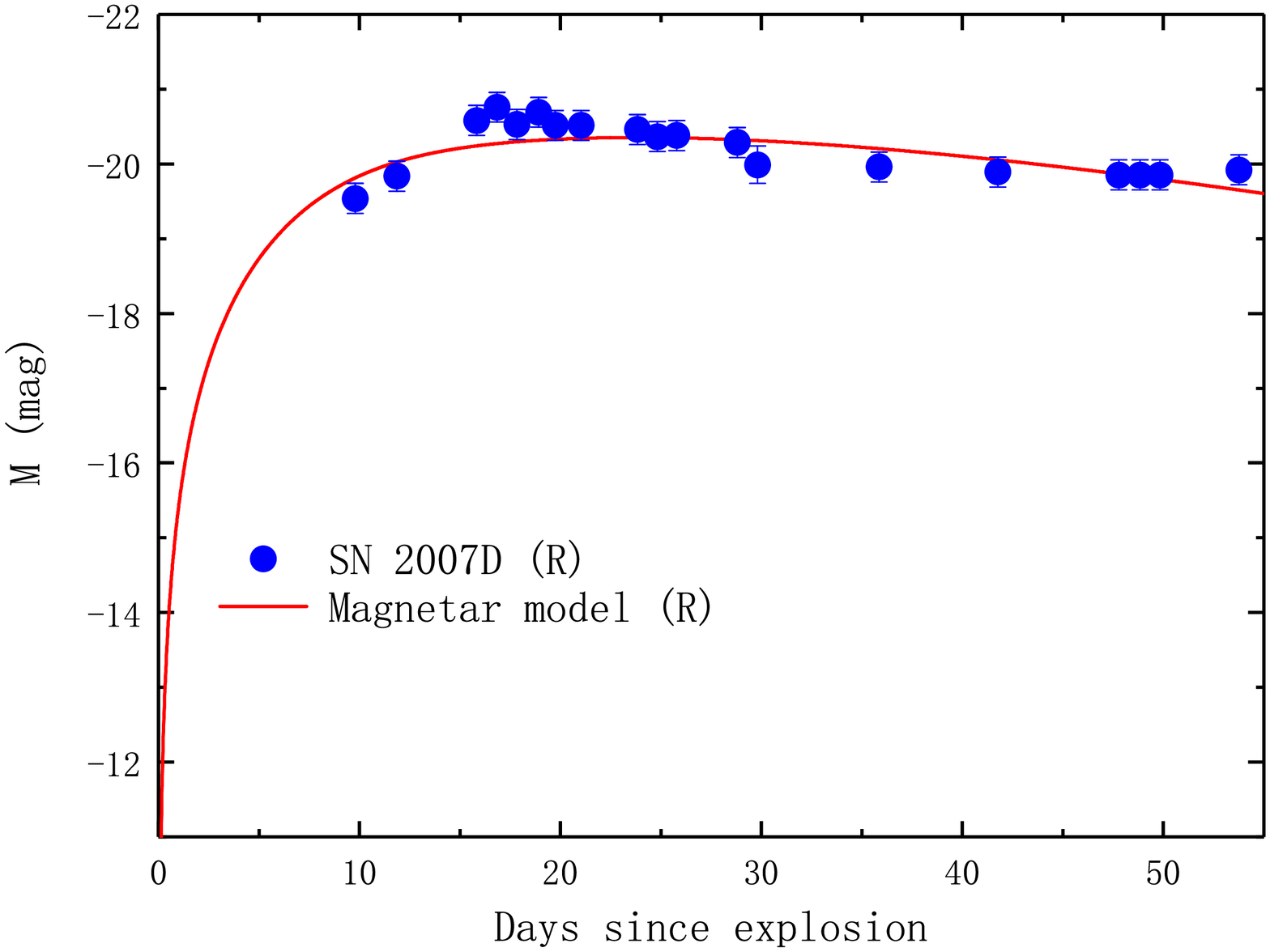}
\includegraphics[width=0.45\textwidth,angle=0]{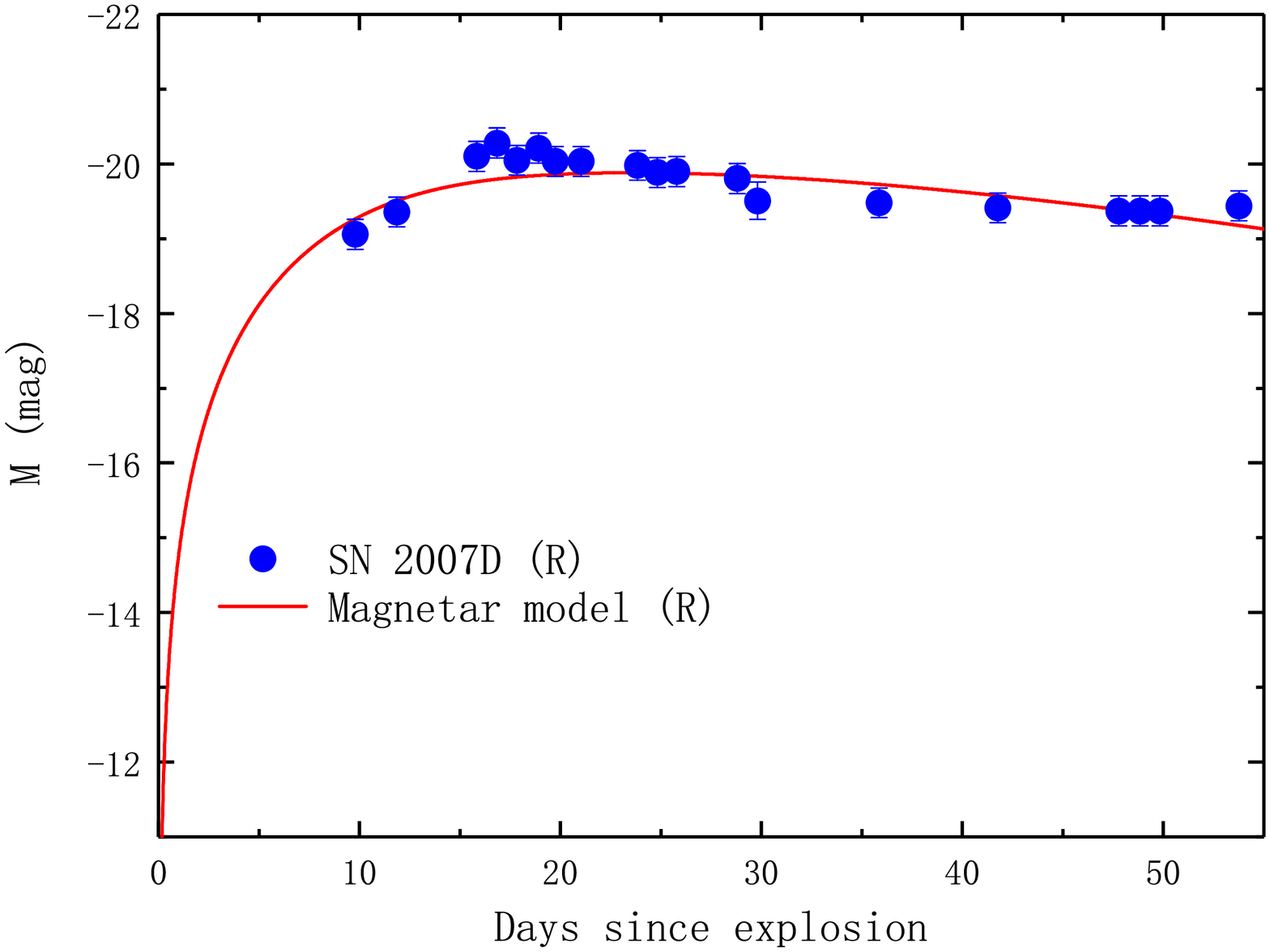}
\includegraphics[width=0.45\textwidth,angle=0]{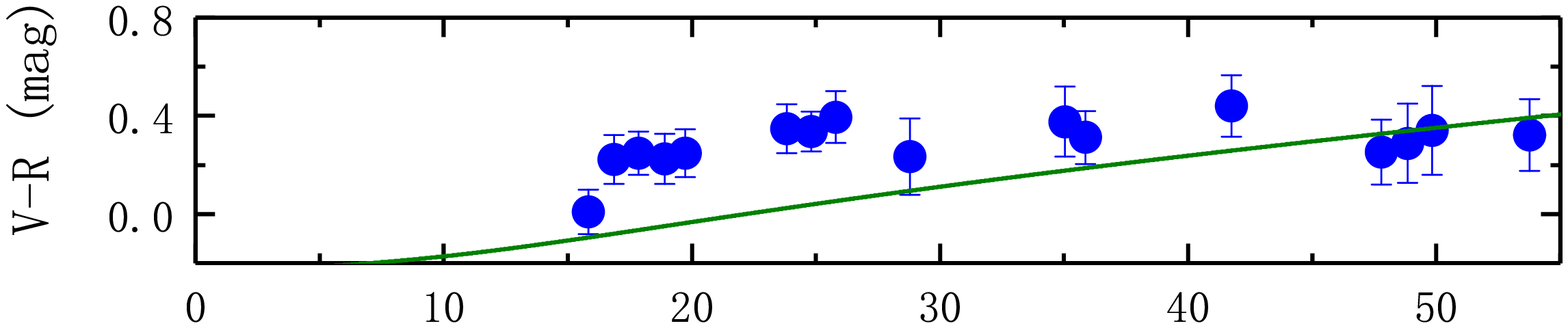}
\includegraphics[width=0.45\textwidth,angle=0]{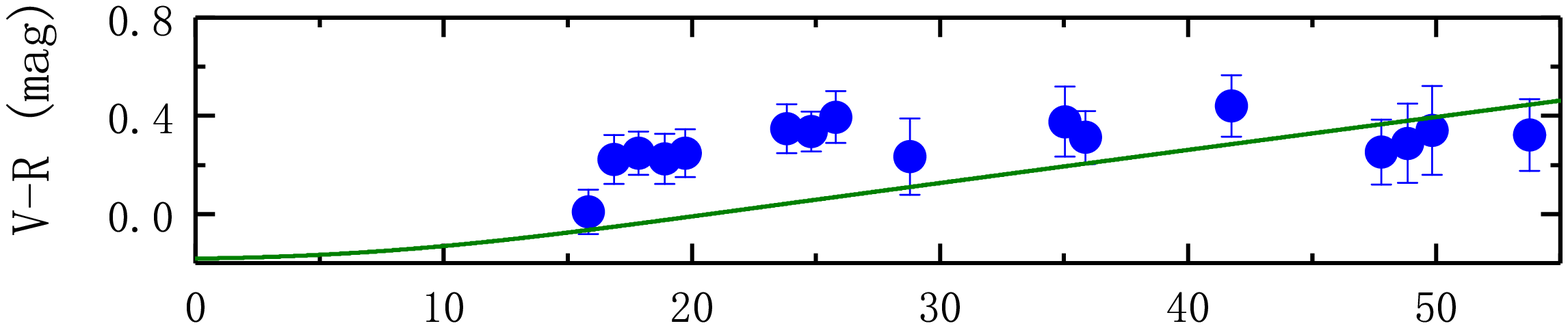}
\end{center}
\caption{The $R$-band LCs (top panels) and the $V-R$ color evolution
(bottom panels) reproduced by the magnetar model for Case A (left) and Case B (right).
Data for Case A are taken from \citet{Dro2011}. The abscissa
represents time since the explosion in the rest frame.}
\label{fig:2007D-mag}
\end{figure}

\clearpage

\begin{figure}[tbph]
\begin{center}
\includegraphics[width=0.45\textwidth,angle=0]{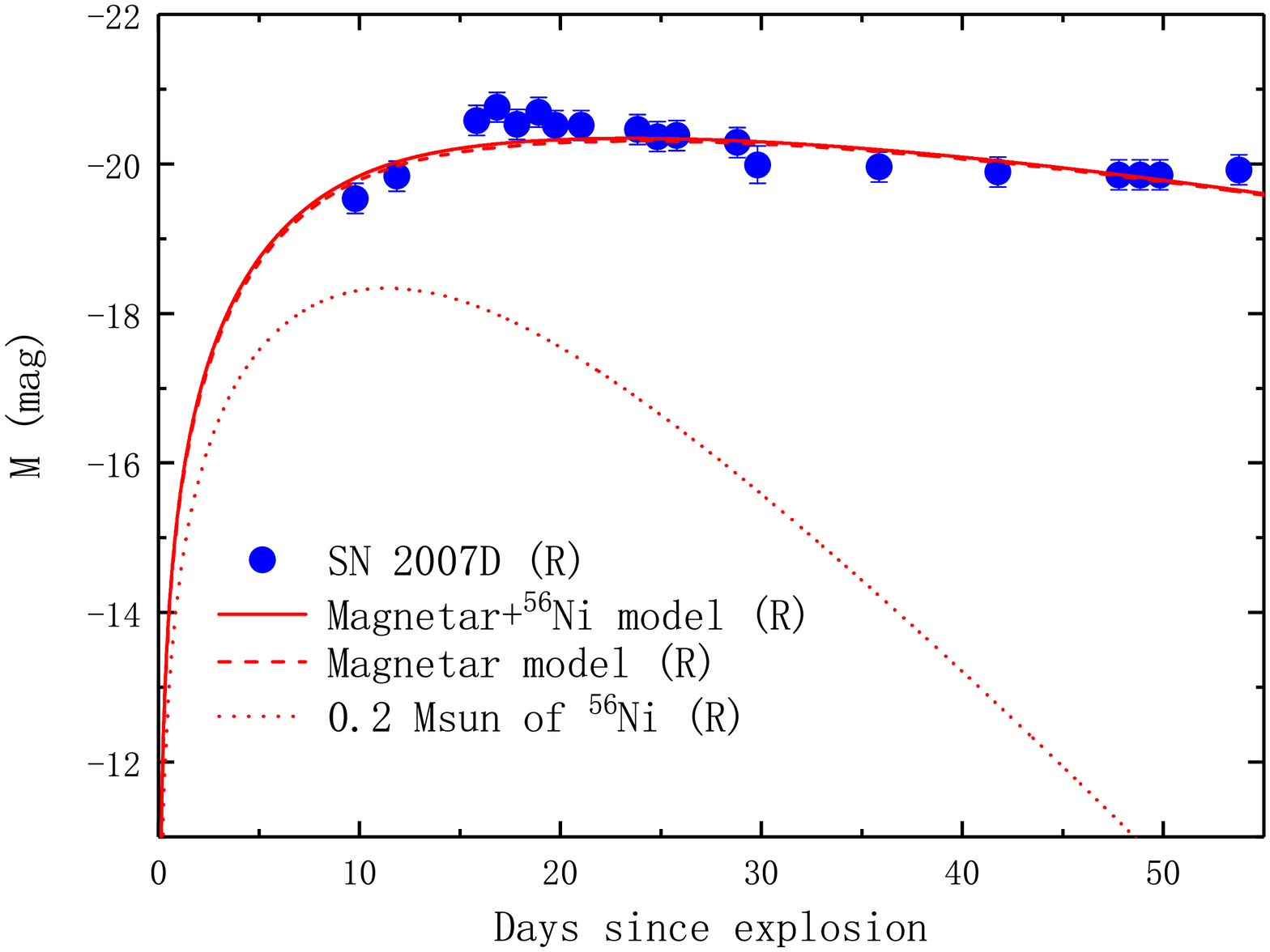}
\includegraphics[width=0.45\textwidth,angle=0]{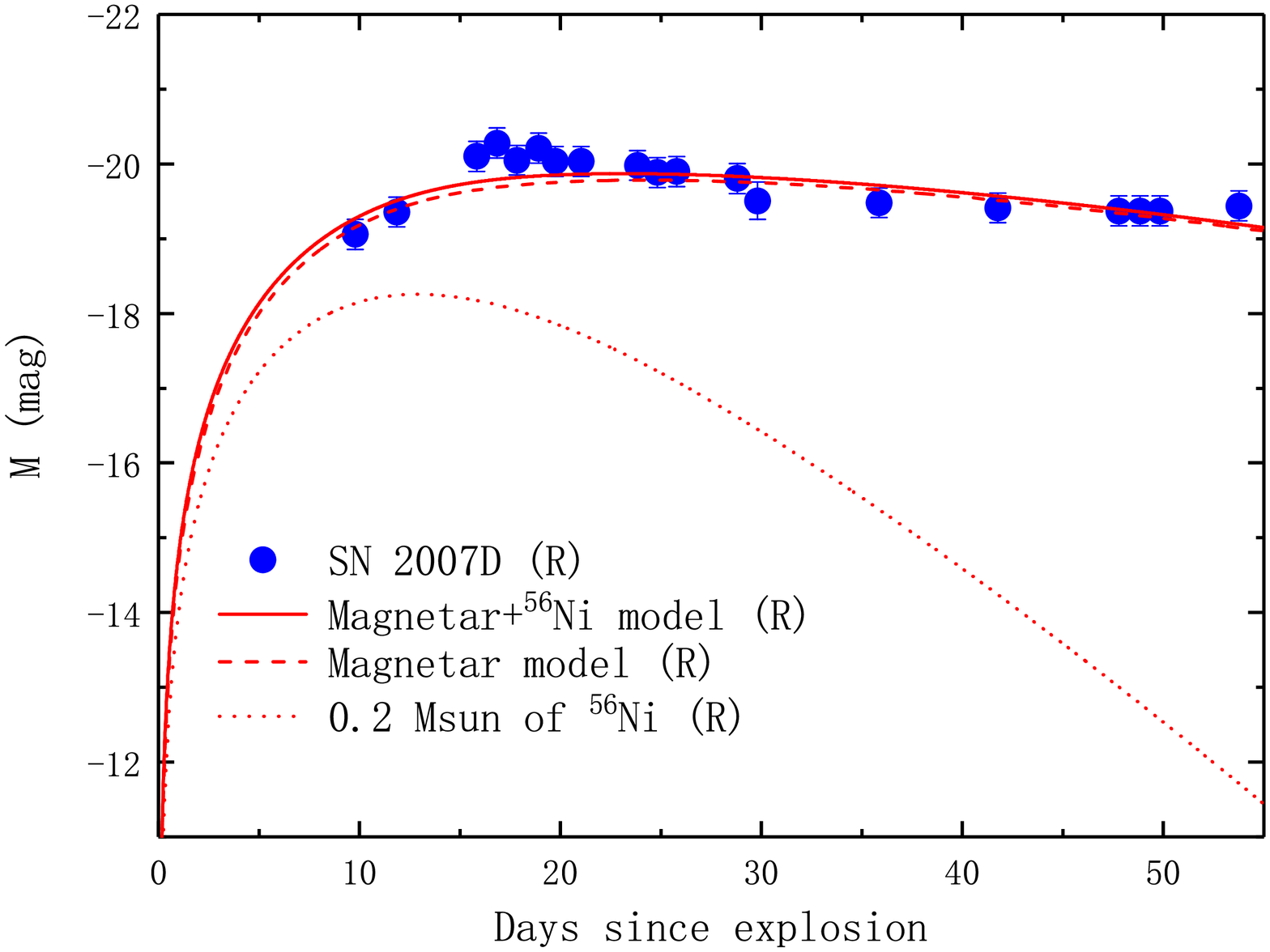}
\includegraphics[width=0.45\textwidth,angle=0]{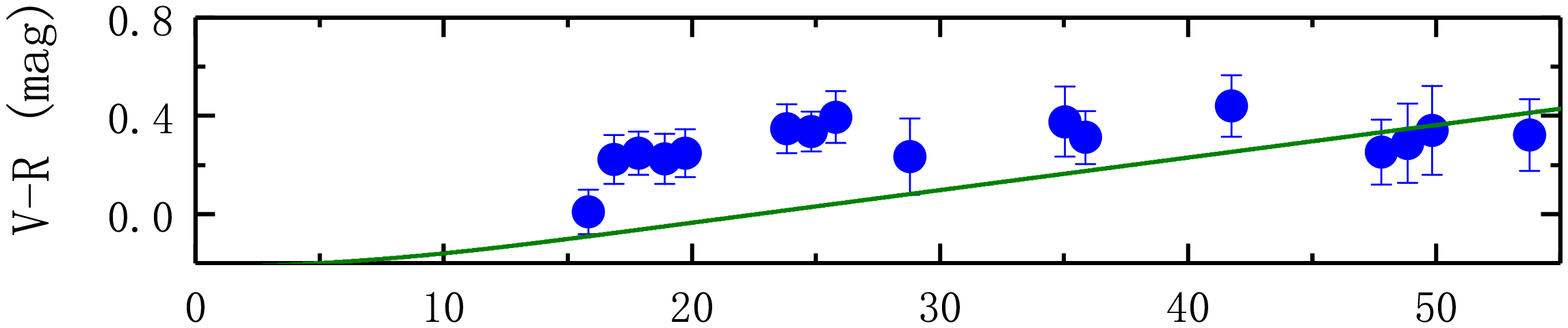}
\includegraphics[width=0.45\textwidth,angle=0]{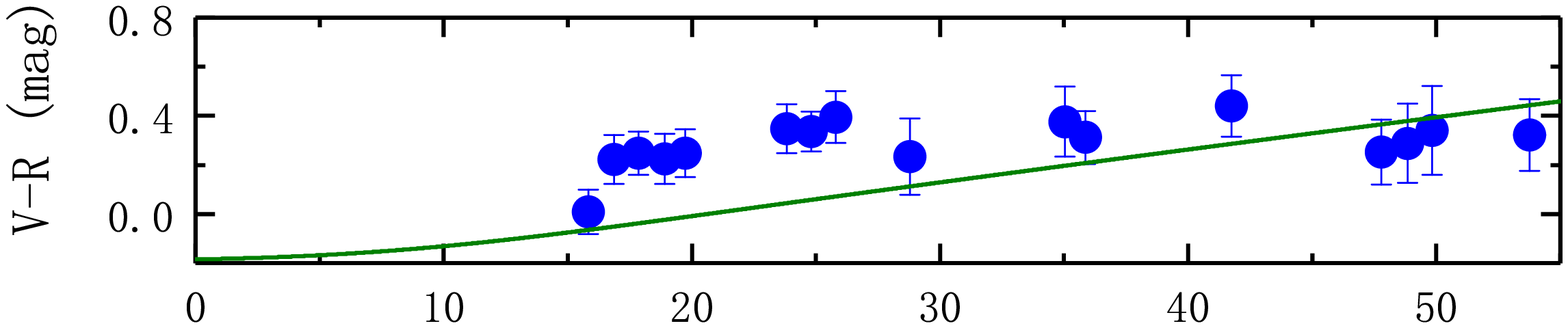}
\end{center}
\caption{The $R$-band LCs (top panels) and the $V-R$ color evolution
(bottom panels) reproduced by the magnetar+$^{56}$Ni model
for Case A (left) and Case B (right). 
The dashed and dotted lines represent
the LCs of the components from magnetar and the 0.2~M$_\odot$ of $^{56}$Ni, respectively.
Data for Case A are taken from \citet{Dro2011}. The abscissa
represents time since the explosion in the rest frame.}
\label{fig:2007D-magni}
\end{figure}

\clearpage

\begin{figure}[tbph]
\begin{center}
\includegraphics[width=0.45\textwidth,angle=0]{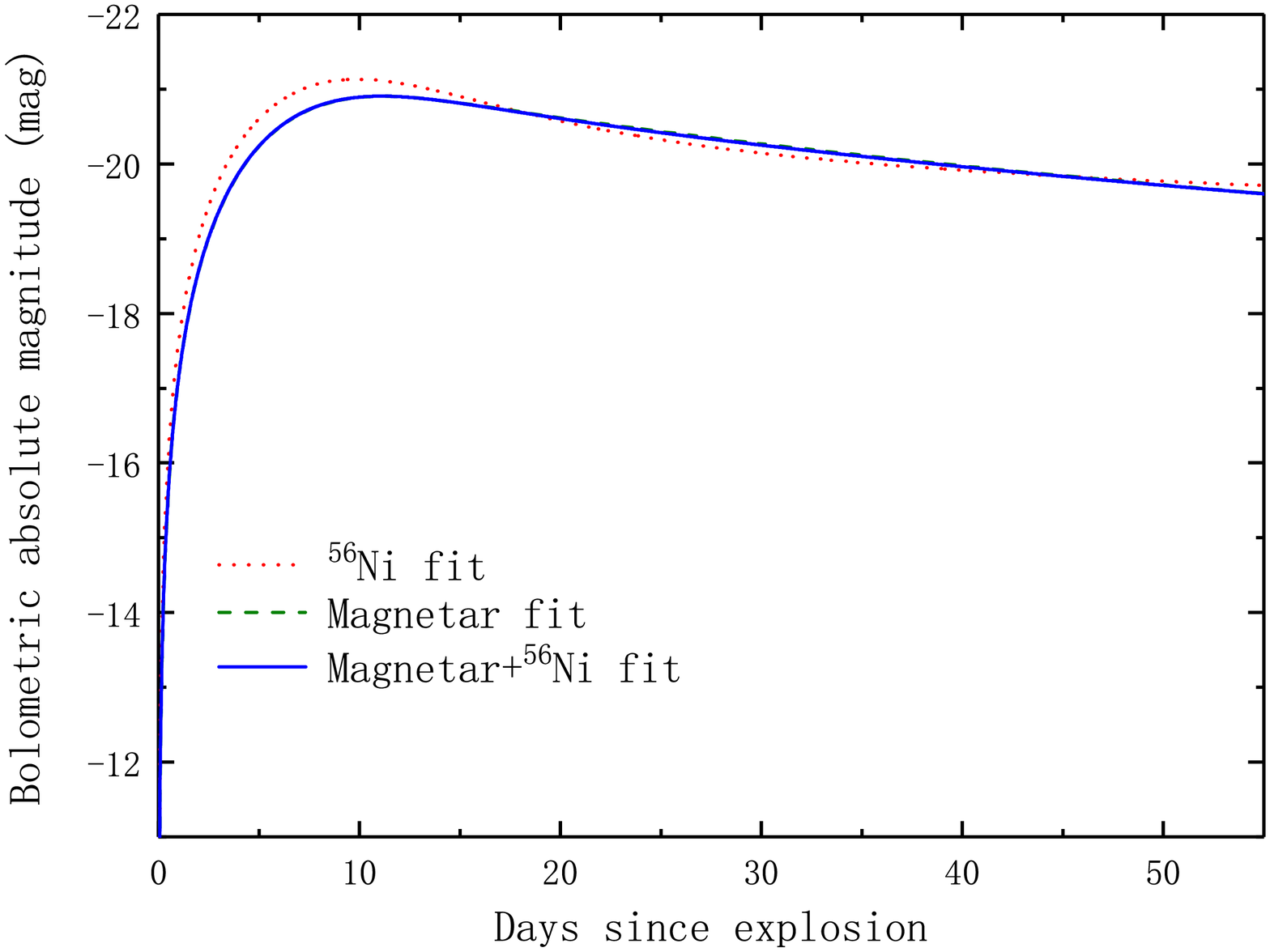}
\includegraphics[width=0.45\textwidth,angle=0]{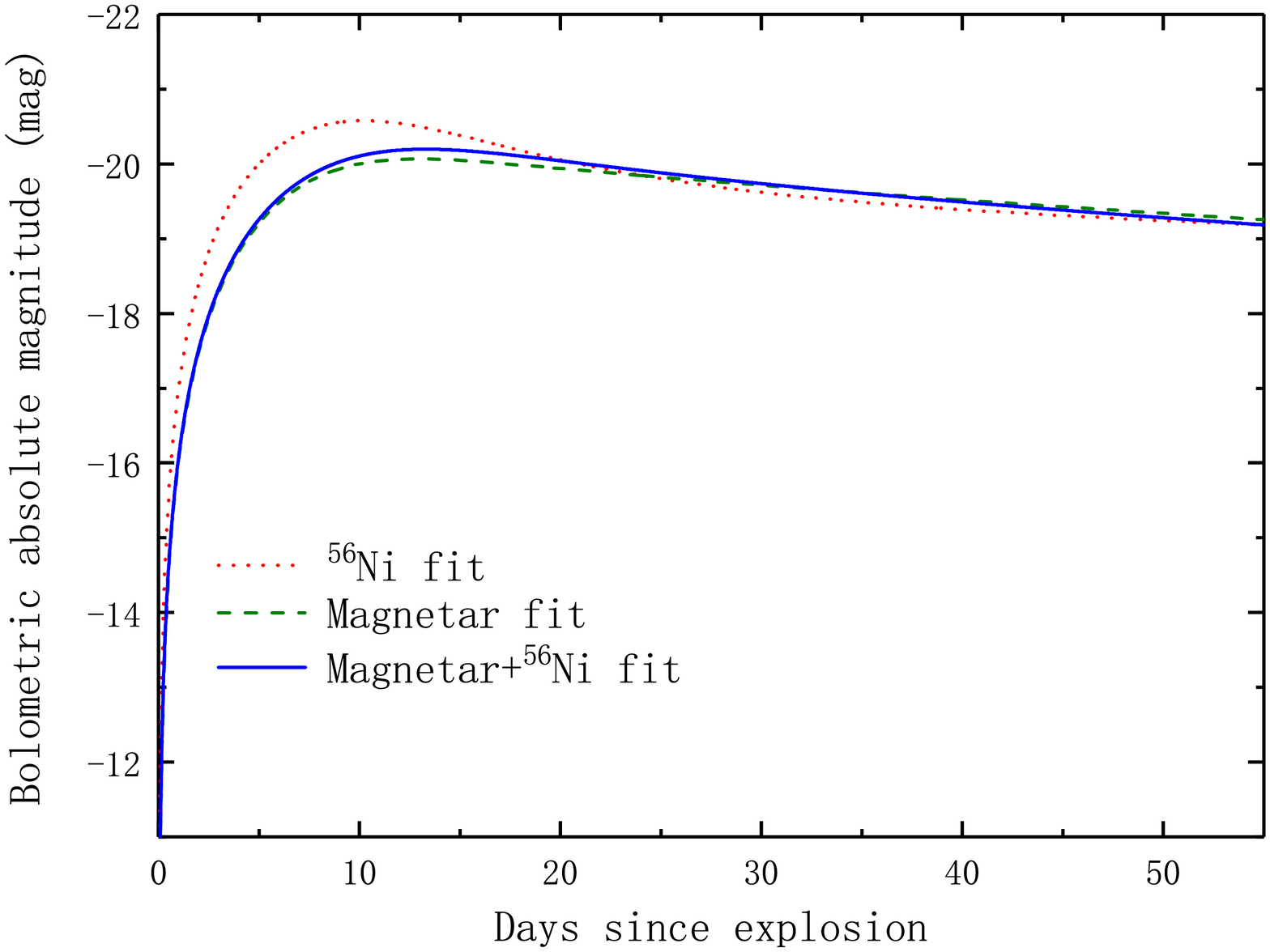}
\includegraphics[width=0.45\textwidth,angle=0]{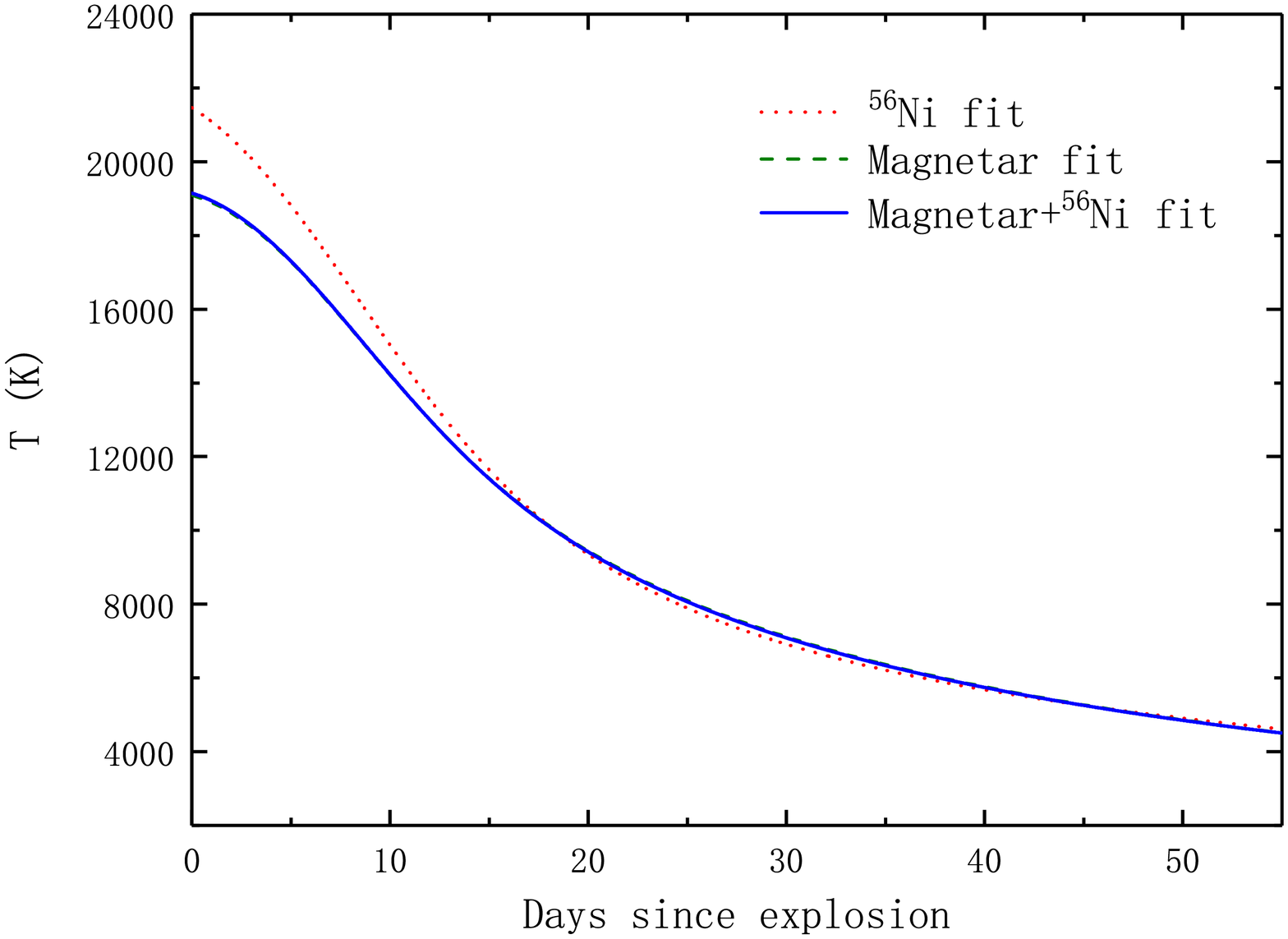}
\includegraphics[width=0.45\textwidth,angle=0]{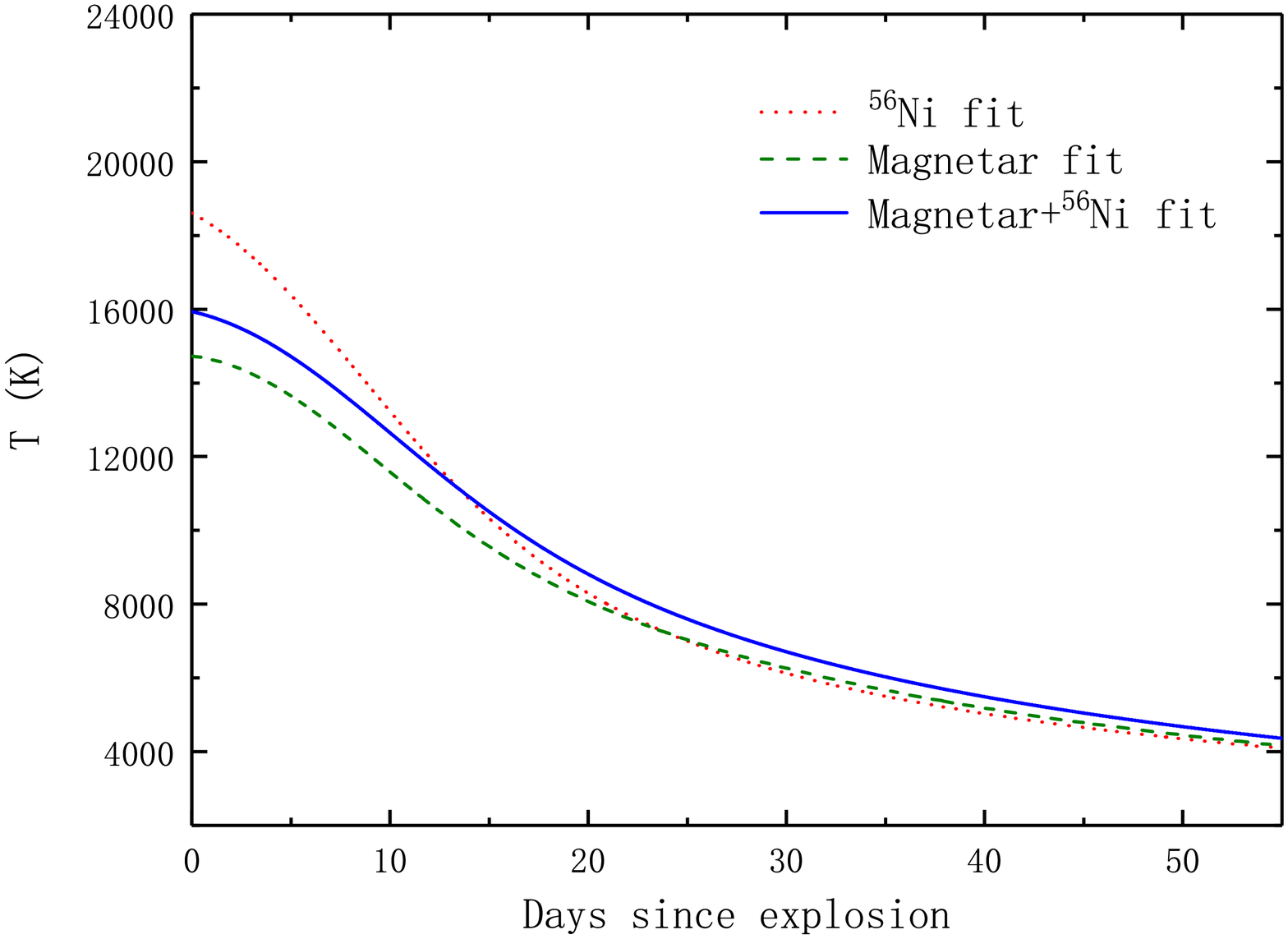}
\end{center}
\caption{The bolometric LCs (top panels) and the temperature evolution
(bottom panels) reproduced by the $^{56}$Ni model, the magnetar model, and
the magnetar+$^{56}$Ni model for Case A (left panels) and Case B (right panels).
The abscissa represents time since the explosion in the rest frame.}
\label{fig:2007D-BoloTV}
\end{figure}

\clearpage

\begin{figure}[tbph]
\begin{center}
\includegraphics[width=0.8\textwidth,angle=0]{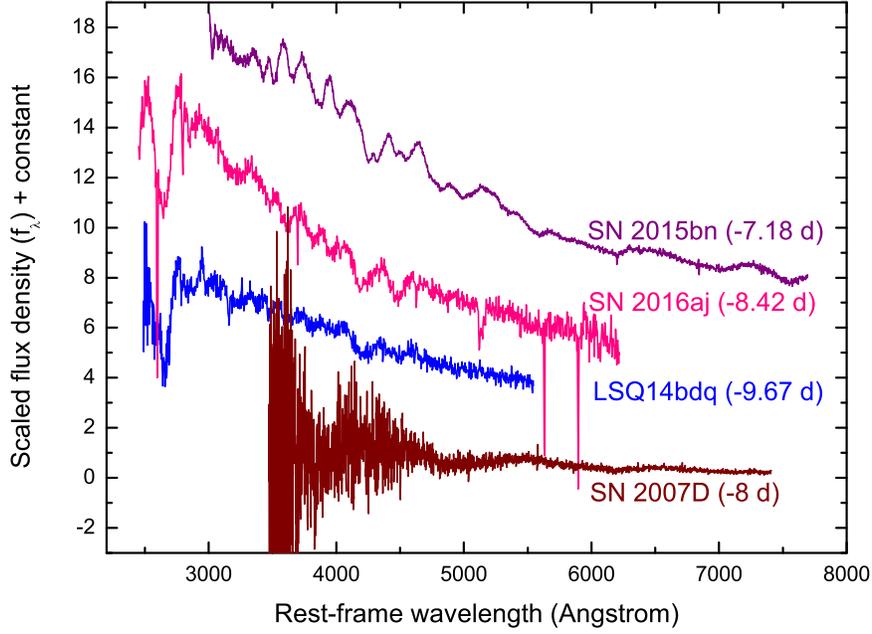}
\end{center}
\caption{Rest-frame premaximum spectra of SN~2007D and three SLSNe-I. 
The spectrum of SN~2007D was obtained from the CfA Supernova Archive \citep{Mod2014} and
corrected by taking the extinction ($E(B-V) = 0.91 + 0.335 = 1.245$~mag) into account.
Spectra of the other three SLSNe (LSQ14bdq, SN~2016aj, and SN~2015bn)
were obtained from 
the Weizmann Interactive Supernova Data Repository (WISeREP) \citep{Yar2012},
the Transient Name Server (https://wis-tns.weizmann.ac.il/), and \citet{Nich2016}, respectively.
The extinction-corrected spectrum of SN~2007D is cooler than that of
these SLSNe-I, indicating that SN~2007D might be a luminous SN rather than a SLSN-I.}
\label{fig:spec}
\end{figure}

\end{document}